\documentclass[12pt,a4paper]{article}
\usepackage{latexsym,amsmath,amssymb,bm, color}

\newtheorem{thm}{Theorem}
\newtheorem{prop}[thm]{Proposition}
\newtheorem{lem}[thm]{Lemma}

\newtheorem{exmp}[thm]{Example}
\newtheorem{rem}[thm]{Remark}
\newtheorem{cor}[thm]{Corollary}
\newtheorem{conj}[thm]{Conjecture}

\newcommand{\pf}{{\bf Proof. \ }}
\newcommand{\qed}{\hfill $\Box$

\medskip}

\begin{document}
\title{On the covering dimension of a linear code}
\author{
Thomas Britz\footnote{School of Mathematics and Statistics,
                      UNSW Australia, Sydney, NSW 2052, Australia
                      (email: {\tt britz@unsw.edu.au})}
\;\:\;and\;
Keisuke Shiromoto\footnote{Department of Mathematics and Engineering,
                      Kumamoto University, Kurokami, Kumamoto 862-8555, Japan
                      (email: {\tt keisuke@kumamoto-u.ac.jp})}}

\maketitle

\begin{abstract}
The critical exponent of a matroid is one of the important parameters in matroid theory
and is related to the Rota and Crapo's Critical Problem.
This paper introduces the covering dimension of a linear code over a finite field,
which is analogous to the critical exponent of a representable matroid.
An upper bound on the covering dimension is conjectured and nearly proven,
improving a classical bound for the critical exponent.
Finally, a construction is given of linear codes
that attain equality in the covering dimension bound.

\end{abstract}

\section{Introduction}
\label{sec:intro}

The Critical Problem of matroid theory, posed by Crapo and Rota~\cite{crapo-rota70},
is to determine for $q$-representable classes of matroids $M$ the invariant $c(M;q)$ known as the {\em critical exponent}.
This problem has attracted much interest; see for instance \cite{AnBo02, AsNiOxSa84, britz05, britz10, dowling71, kung00, lindstrom78, oxley78, whittle84, whittle89}
and the excellent monograph-length survey~\cite{kung96} on this problem and variations thereof.
One reason for this interest is that the critical exponent generalises the chromatic polynomial for graphs
and, it was once hoped, methods for evaluating critical exponent could lead to a short and transparent proof of
the famous Four Colour Theorem, which at that time was not proved to everyone's satisfaction.

These hopes have so far been frustrated since the critical exponent has turned out to be very difficult to determine,
and the Critical Problem has turned out to be significantly more challenging - and of fundamental interest -
than the Four Colour Problem.
Until now, there are only very few general results to describe the critical exponent,
one of which is the following bound by Kung~\cite[(4.10)]{kung96}:
%[Kung's Propositions are now Theorems: they are the focus of our paper,
%so it might be a good idea to make them more prominent in this way.]

\medskip

\begin{thm}
\label{thm:kung96}
If $M$ is a rank $k$ simple matroid that is representable over $\mathbb{F}_q$, then
\[
  c(M;q) \leq k-g+3\,,
\]
where $g$ denotes the girth of $M$, that is, the minimum cardinality of circuits of $M$.
\end{thm}
%{\bf [Preceding (4.10) is (4.24), a lemma that is a bit stronger than (4.10) and has a simple coding-theoretical proof. Is this perhaps useful?]}

\medskip

The purpose of this paper is to introduce the {\em covering dimension} for linear codes over a finite field $\mathbb{F}_q$,
analogous to the critical exponent for representable matroids.
In Section~\ref{sec:betterbound},
we conjecture an upper bound on the covering dimension (cf.\ Conjecture~\ref{conj:conjecture})
that improves Kung's bound above,
and we nearly prove this conjecture by
showing that the upper bound is true for the great majority of linear codes (cf.\ Corollary~\ref{cor:mostcases}).
Furthermore, in Section~\ref{sec:construction},
we provide constructions of linear codes over finite fields that attain our improved bound,
thus proving that it is tight.

%%%%%%%%%%%%%%%%%%%%
\section{Preliminaries}
Set $E:=\{1,\ldots,n\}$.
Let $\mathbb{F}_q^n:=\mathbb{F}_q^E$ be the vector space of ordered $n$-tuples of elements from $\mathbb{F}_q$ indexed by $E$.
The {\em support} and {\em weight} of each vector $\bm{x}=(x_1,\ldots,x_n) \in \mathbb{F}_q^n$ is given by
\begin{align*}
  \textrm{supp}(\bm{x}) &:= \{\,i \:\::\:\: x_i \neq 0\}\,;\\
    \textrm{wt}(\bm{x}) &:= |\textrm{supp}(\bm{x})|\,.
\end{align*}
Similarly, the {\em support} and {\em weight} of each subset $B\subseteq \mathbb{F}_q^n$ are defined as follows:
\begin{align*}
  \textrm{Supp}(B)  &:= \bigcup_{\bm{x} \in B} \textrm{supp}(\bm{x})\,;\\
  \textrm{wt}(B)    &:= |\,\textrm{Supp}(B) |\,.
\end{align*}
For vectors $\bm{x} = (x_1,\ldots,x_n), \bm{y} = (y_1,\ldots,y_n)\in \mathbb{F}_q^n$, define the {\em inner product}
\[
  \bm{x}\cdot\bm{y} := \sum_{i=1}^nx_iy_i\,.
\]
Let $C$ be an $[n,k]$ code $C$ over $\mathbb{F}_q$;
that is, a $k$-dimensional subspace of the vector space $\mathbb{F}_q^n$.
%and let $G$ be a $k \times n$ (generator) matrix over $\mathbb{F}_q$ whose rows form a basis for~$C$.
The {\em dual code} $C^{\perp}$ of $C$ is defined by
\[
  C^{\perp} := \{\bm{y} \in \mathbb{F}_q^n \::\: \bm{x}\cdot\bm{y} = 0 \: \textrm{ for all } \bm{x} \in C\}\,.
\]
The {\em minimum Hamming weight} of $C$ is defined by
\[
  d := d(C) = \min\{\textrm{wt}(\bm{x}) \::\: (\bm{0} \neq )\bm{x} \in C \}\,.
\]
The minimum Hamming weight of the dual code $C^{\perp}$ is often simply denoted by $d^\perp:=d(C^\perp)$.
An $[n,k,d]$ code $C$ is an $[n,k]$ code with minimum distance $d(C) = d$.

We now introduce the {\em covering dimension} of $C$, denoted by $\gamma(C)$,
as follows:
%For any $r$, $0 \le r \leq k$ and any $i$, $1\leq i \leq n$,
%let $A_i^{(r)} := A_i^{(r)}(C)$ be the number of $r$-dimensional subcodes $D$ of $C$ with $\textrm{wt}(D) = i$.
%The polynomial
%\[
%  W_C^{(r)}(x,y) = \sum_{i=0}^nA_i^{(r)}x^{n-i}y^i
%\]
%is the {\em $r$th support weight enumerator} of $C$ (cf.~\cite{klove92}).\\
%We define the {\em covering dimension} of $C$, denoted by $\gamma(C)$,
\[
  \gamma(C) :=
    \begin{cases}
      \infty                                                 &,\; \textrm{if} \; \textrm{Supp}(C) \neq E\,;\\
      \min\{r \in \mathbb{Z}^{+} \::\: A_n^{(r)}(C) \neq 0\} &,\; \textrm{otherwise}
    \end{cases}
\]
where
%For any $r$, $0 \le r \leq k$ and any $i$, $1\leq i \leq n$,
%let $A_i^{(r)} := A_i^{(r)}(C)$ be the number of $r$-dimensional subcodes $D$ of $C$ with $\textrm{wt}(D) = i$.
$A_i^{(r)}(C)$ is the number of $r$-dimensional subcodes $D$ of $C$ with $\textrm{wt}(D) = i$.
% or, equivalently, $\textrm{Supp}(D) = E$.

We will now briefly show the connection between the covering dimension $\gamma(C)$ and the critical exponent $c(M;q)$ for matroids.

For each subset $X \subseteq E$, the {\em shortened code}, denoted by $C/X$,
is the linear code obtained by deleting the (zero) coordinates $X$ from
each codewords $\bm{x} \in C$ with $\textrm{ supp}(\bm{x}) \cap X = \emptyset$.
The {\em punctured code}, denoted by $C \setminus X$,
is the linear code obtained by deleting the coordinate~$X$ from each codeword in~$C$.
The function $\rho$ given by $\rho(X):=\dim C\setminus (E-X)$ for $X\subseteq E$ induces a matroid $M_C=(E,\rho)$
(we refer the reader to~\cite{oxley92, welsh76} for information on matroids).
%Conversely, if $M$ is a matroid representable over~$\mathbb{F}_q$,
%then a linear code $C$ exists such that $M = M_C$
%Also it is well known that the circuits of $M_C$ are precisely
%the minimal nonempty codeword supports of the dual code $C^{\perp}$ (cf.~\cite{oxley92}).
The {\em critical exponent} $c(M_C;q)$ of $M_C$ is defined by
\[
  c(M_C;q)=
  \begin{cases}
    \infty                                          &,\;\textrm{if $M_{C}$ has a loop}; \\
    \min\{j \in \mathbb{Z}^{+} \::\: p(M_C;q^j)>0\} &,\;\textrm{otherwise}
  \end{cases}
\]
where $p(M_C;q^j)$ is the {\em characteristic polynomial} of $M_C$,
defined by
\[
  p(M_C;\lambda)=\sum_{X \subseteq E}(-1)^{|X|}\lambda^{\rho(E)-\rho(X)}\,.
\]
%Let $M$ be a matroid representable over $GF(q) = \mathbb{F}_q$.
%It is well known that $p(M;q^k) \geq 0$ for all $k \in \mathbb{Z}^{+}$.
%
%Thus if $M$ has no loops, then $p(M;q^k)>0$ for all $k \geq c(M;q)$.
%For a matroid $M$ which is representable over $\mathbb{F}_q$,
The Critical Problem of matroid theory is to
%one of the so-called Critical Problems in matroid theory is the problem of
determine $c(M_C;q)$
(cf.~\cite{AsNiOxSa84, britz05, dowling71, kung96}).
One of the main tools that has been used to address the Critical Problem
is the following result widely known as the {\em Critical Theorem}
by Crapo and Rota~\cite{crapo-rota70} (see also \cite[Theorem~2]{britz05}).

\medskip

\begin{thm}
\label{thm:critical-theorem}
{\rm (The Critical Theorem)}
Let $C$ be an $[n,k]$ code over $\mathbb{F}_q$.
For any $X \subseteq E$ and any $m \in \mathbb{Z}^{+}$,
the number of ordered $m$-tuples $(\bm{v}_1,\ldots,\bm{v}_m)$ of codewords $\bm{v}_1,\ldots,\bm{v}_m$ in $C$
with $\textrm{supp}(\bm{v}_1)\cup\cdots\cup\textrm{supp}(\bm{v}_m) = X$ is $p(M_{C/(E-X)}; q^m)$.
\end{thm}

\medskip

By applying $E = X$ to the Critical Theorem,
we can equate the covering dimension $\gamma(C)$ of a linear code
over $\mathbb{F}_q$ and the critical exponent $c(M_C;q)$ of its induced matroid~$M_C$.

\medskip

\begin{lem}
\label{lem:equiv}\:\:
$\gamma(C) = c(M_C;q)$.
\end{lem}

\medskip

\pf
If $c(M_C;q) = m$, then the Critical Theorem implies that
there is at least one set $V = \{\bm{v}_1,\ldots,\bm{v}_m\}$ of codewords in $C$ with $\textrm{Supp}(V) = E$.
Hence, $\gamma(C) \leq m = c(M_C;q)$.
Conversely, suppose that $\gamma(C) = m'$.
By definition, at least one $m'$-dimensional subspace $V$ generated by $m'$ codewords in $C$ satisfies $\textrm{Supp}(V)=E$.
Hence, $c(M_C;q) \leq m' = \gamma(C)$.
\qed

\medskip

Lemma~\ref{lem:equiv} implies that the covering dimension $\gamma(C)$ of a linear code
is fundamentally of a set-theoretic and combinatorial nature,
rather than being strictly algebraic.
Conversely, the lemma lends the important observation that the critical
exponent of a representable matroid $M$ does not depend on any particular linear code $C$ for which $M=M_C$.
As a corollary to Lemma~\ref{lem:equiv}, and using the fact
that the circuits of $M_C$ are also the minimal nonempty codeword supports of
the dual code $C^{\perp}$ (cf.~\cite{oxley92}),
we may re-cast Theorem~\ref{thm:kung96} as follows:

\medskip

\begin{thm}
\label{thm:kung96-code}
If $C$ is an $[n,k]$ code over $\mathbb{F}_q$ with $d^{\perp} := d(C^\perp)\geq 3$,
then
\begin{eqnarray}\label{bound:kung-code}
  \gamma (C) \leq k-d^{\perp}+3\,.
\end{eqnarray}
\end{thm}

Let $A$ be an $[n,r]$ subcode of $C$ with $\textrm{Supp}(A)=E$ having generator matrix~$MG$
where $M$ is an $r \times k$ matrix over~$\mathbb{F}_q$.
Let $P$ be the $[k,r]$ code over $\mathbb{F}_q$ having generator matrix $M$.
Then the dual code $P^{\perp}$ does not contain any of the $n$ column vectors of~$G$.

Conversely, let $U$ be a $(k-r)$-dimensional subspace of $\mathbb{F}_q^k$
which does not contain any of the $n$ column vectors of~$G$.
Let $M$ be a generator matrix of the dual code $U^{\perp}$.
Then the code $S$ having generator matrix $MG$ is an $[n,r]$ subcode of $C$ with $\textrm{Supp}(S) = E$.
Therefore we have the following correspondence (cf.~\cite[Lemma~2]{klove92}):

\medskip

\begin{lem}
\label{lem:one-to-one}
There exists an $[n,r]$ subcode $A \leq C$ with $\textrm{Supp}(A)=E$ if and only if
there is a $(k-r)$-dimensional subspace $U$ of $\mathbb{F}_q^k$
which contains none of the $n$ column vectors of~$G$.
\end{lem}

\medskip

We summarize the above results as follows:

\medskip

\begin{prop}
\label{prop:summary}
Let $C$ be an $[n,k]$ code over $\mathbb{F}_q$ with generator matrix~$G$.\\
The following are equivalent:
\begin{enumerate}
\item[{\rm (1)}] $c(M_C;q)  = m$;
\item[{\rm (2)}] $\gamma(C) = m$;
\item[{\rm (3)}] $m$ is the smallest integer for which
  a $(k-m)$-dimensional subspace $U$ of $\mathbb{F}_q^k$ exists
  that contains no column vector of~$G$.
\end{enumerate}
\end{prop}

%%%%%%%%%%%%%%%%%%%%
\section{A modified bound}
\label{sec:betterbound}

\subsection{Codes attaining Kung's bound}
Let $G$ be a $k \times n$ matrix over $\mathbb{F}_q$
which contains as columns exactly one multiple of each nonzero vector in $\mathbb{F}_q^k$.
Then the $[n=(q^k-1)/(q-1),k]$ code $\mathcal{H}^{\perp}$ with generator matrix $G$ is
a {\em dual Hamming code} and $(\mathcal{H}^{\perp})^{\perp}$ is a $[n,n-k,3]$ Hamming code.

It is also known that, for each $r = 1,\ldots,k$,
\[
    A_{i}^{(r)}(\mathcal{H}^{\perp})
  = \begin{cases}
      \genfrac{[}{]}{0pt}{}{k}{r}_q & ,\; \textrm{if} \; i = (q^k-q^{k-r})/(q-1);\\
       0                            & ,\; \textrm{otherwise}
    \end{cases}
\]
where $\genfrac{[}{]}{0pt}{}{k}{r}_q$ denotes the Gaussian binomial coefficient (cf.~\cite{klove92}).
We see that $i = n$ if and only if $r = k$.
Therefore, $\gamma(\mathcal{H}^{\perp}) = k \,(= k-3+3)$,
and $\mathcal{H}^{\perp}$ attains the bound in Theorem~\ref{thm:kung96-code}.

A {\em maximum distance separable} (MDS) code over $\mathbb{F}_q$ is an $[n,k]$ code over $\mathbb{F}_q$
whose minimum Hamming weight is $n-k+1$.
Since the dual code of an MDS $[n,k]$ code over $\mathbb{F}_q$ is also an MDS code,
it follows that an $[n,k]$ code $C$ is an MDS code if and only if $d^{\perp} = k+1$.
According to~\cite[Theorem~6, p.~321]{ms77},
the number $A_w$ of codewords of weight $w$ in an MDS $[n,k]$ code over $\mathbb{F}_q$ is given by
\begin{eqnarray}
\label{eqn:weightdistributions}
  A_w = \binom{n}{w}(q-1)\sum_{j=0}^{w-d}(-1)^j\binom{w-1}{j}q^{w-d-j}\,,
\end{eqnarray}
for $d\leq w \leq n$, where $d=n-k+1$.

From Equation~(\ref{eqn:weightdistributions}), we have that
\begin{eqnarray*}
       \frac{A_n}{q-1}
  &=&  \sum_{j=0}^{k-1}(-1)^j\binom{n-1}{j}q^{k-1-j}\\
  &=&  \binom{n-1}{0}q^{k-1}+\sum_{j=1}^{k-1}(-1)^j\left[\binom{n-2}{j-1}+\binom{n-2}{j}\right]q^{k-1-j}\\
  &=&  \sum_{j=1}^{k-1}(-1)^{j}\binom{n-2}{j-1}q^{k-1-j}+ \sum_{j=0}^{k-1}(-1)^j\binom{n-2}{j}q^{k-1-j}\\
  &=& -\sum_{j=0}^{k-2}(-1)^{j}\binom{n-2}{j  }q^{k-2-j}+q\sum_{j=0}^{k-2}(-1)^j\binom{n-2}{j}q^{k-2-j}+(-1)^{k-1}\binom{n-2}{k-1}\\
  &=& -\frac{A_{n-1}}{n(q-1)}+\frac{qA_{n-1}}{n(q-1)}+(-1)^{k-1}\binom{n-2}{k-1}\\
  &=&  \frac{A_{n-1}}{n}+(-1)^{k-1}\binom{n-2}{k-1}.
\end{eqnarray*}
Therefore, if $A_n=0$, then $A_{n-1}=(-1)^{k}n\binom{n-2}{k-1}\neq 0$ and $k$ is even,
and so $\gamma(C)=2\,(=k-(k+1)+3)$.

We summarize the above results as follows:

\medskip

\begin{prop}
\label{prop:kung's codes}
If an $[n,k]$ code $C$ over $\mathbb{F}_q$ is a dual Hamming code or an MDS code
having no codewords of weight $n$, then $C$ attains the bound in Theorem~\ref{thm:kung96-code}.
\end{prop}

\medskip

\begin{rem}
\label{rem:MDS}
{\rm
In \cite{MDS}, it is shown that
if $C$ is an MDS code over $\mathbb{F}_q$ of length $n\leq q+2$ with no codewords of weight $n$,
then $C$ is a binary $[n,n-1]$ code with odd $n$ or a $[q+1,2]$ (dual Hamming) code over $\mathbb{F}_q$.}
\end{rem}

\medskip

In the following propositions,
we classify two important classes of codes that attain
the bound in Theorem~\ref{thm:kung96-code},
i.e., Kung's bound for codes.
These propositions will be used in the following section.

\medskip

\begin{prop}
\label{prop:d=3}
Let $C$ be a linear $[n,k]$ code over $\mathbb{F}_q$ with $d^{\perp}=3$.
Then
$$\gamma(C) = k-d^{\perp}+3\; (=k-3+3=k)$$
if and only if $C$ is isomorphic to a dual Hamming code.
\end{prop}

\medskip

\pf
From Proposition~\ref{prop:summary},
$\gamma(C) = k$ if and only if each $1$-dimensional subspace of $\mathbb{F}_q^k$
contains one column vector of a generator matrix for~$C$;
since $d^\perp = 3$, any two such column vectors are linearly independent,
so $C$ is isomorphic to the dual Hamming code.
\qed

\medskip

\begin{prop}
\label{prop:binaryMDS}
Let $C$ be a binary linear $[n,n-1]$ code.
Then
$$\gamma(C) = (n-1)-d^{\perp}+3\; (=(n-1)-n+3=2)$$
if and only if $n$ is odd.
\end{prop}

\medskip

\pf
$C$ is isomorphic to a binary code $C'$ having generator matrix of the form
\[
  G = \begin{pmatrix}
                & 1 \\[-1.5mm]
        I_{n-1} & \vdots\\
                & 1
      \end{pmatrix}\,.
\]
Therefore,
$C'$ contains $(1,\ldots,1,1+\cdots+1)=(1,\ldots,1,n-1)\: (\mbox{\rm mod} \:\:2)$ as a codeword
, so $\bm{1}=(1,\ldots, 1)\notin C$ if and only if $n$ is odd.
\qed

%%%%%%%%%%%%%%
%%%%%%%%%%%%%%%%
%%%%%%%%%%%%%%%%%%
%%%%%%%%%%%%%%%%%%%%
\subsection{A modified bound}

We now turn to the main aim of the paper
which is to sharpen Kung's bound for codes (Theorem~\ref{thm:kung96-code}).
In light of Proposition~\ref{prop:kung's codes} and inspired by Remark~\ref{rem:MDS},
we conjecture that Kung's bound may be sharpened as follows.

\medskip

\begin{conj}
\label{conj:conjecture}
If $C$ is an $[n,k]$ code over $\mathbb{F}_q$ with $d^{\perp} := d(C^\perp)\geq 3$,
then
\[
  \gamma(C) \leq k-d^{\perp} + 2
\]
unless $C$ is isomorphic to a dual Hamming code
or     $C$ is a binary $[n,n-1]$ code such that $d^{\perp}=n$ is odd,
in either which case
$\gamma(C) = k-d^{\perp}+3$.
\end{conj}

\medskip

This section serves to mostly verify this conjecture.
We first require two auxiliary lemmas.

\medskip

\begin{lem}
\label{lem:independent}
If $\{\bm{u}_1,\ldots,\bm{u}_t\}\subseteq \mathbb{F}_q^m$ is linearly independent,
then $\{\bm{u}_1+\alpha\bm{u}_t,\ldots,\bm{u}_{t-1}+\alpha\bm{u}_t\}$ is also linearly independent
for any $\alpha\in \mathbb{F}_q$.
\end{lem}

\medskip

\pf
The only scalars in $\mathbb{F}_q$ that satisfy
\begin{eqnarray*}
                         && a_1(\bm{u}_1+\alpha\bm{u}_t)+\cdots+a_{t-1}(\bm{u}_{t-1}+\alpha\bm{u}_t)=\bm{0}\,,\\
\text{or, equivalently,} && a_1 \bm{u}_1 +\cdots+a_{t-1}\bm{u}_{t-1}+\biggl(\alpha\sum_{j=1}^{t-1}a_j\biggr)\bm{u}_{t} = \bm{0}\hspace*{3cm}
\end{eqnarray*}
are $a_1=\cdots=a_{t-1}=0$.
\qed

\medskip

\begin{lem}
\label{lem:subspaces}
Let $\{\bm{u}_1,\ldots,\bm{u}_t\}\subseteq \mathbb{F}_q^m$ be linearly independent
and let $D$, $D'$, and $D_0$ be subspaces generated by
$\bm{u}_1+\alpha\bm{u}_t,\ldots,\bm{u}_{t-1}+\alpha\bm{u}_t$,
$\bm{u}_1+\beta \bm{u}_t,\ldots,\bm{u}_{t-1}+\beta \bm{u}_t$, and
$\bm{u}_1,\ldots,\bm{u}_{t-1}$, respectively, where $\alpha,\beta \in \mathbb{F}_q$.
Then $D=D'$ if and only if $\alpha=\beta$;
furthermore, if $\alpha \neq \beta$, then $D \cap D'\subseteq D_0$.
%\begin{enumerate}
%  \item[{\rm (1)}] $D=D'$ if and only if $\alpha=\beta$;
%  \item[{\rm (2)}] if $\alpha \neq \beta$, then $D \cap D'=\{\bm{0}\}$.
%  %if there exists a vector $\bm{v} \in D$ such that $\bm{v}$ is a column vector of $G$, then $\bm{v} \notin D'$.
%\end{enumerate}
%
% {\bf (2) is not correct: for instance, consider the three standard basis vectors of $\mathbb{F}_3^3$
% and label them $\bm{u}_1$, $\bm{u}_2$, and $\bm{u}_t$. Letting $\alpha = a_1 = 1$ and $\beta = a_2 = -1$,
% we see that $a_1(\bm{u}_1+\alpha\bm{u}_t)+a_2(\bm{u}_2+\alpha\bm{u}_t) = a_1(\bm{u}_1+\beta\bm{u}_t)+a_2(\bm{u}_2+\beta\bm{u}_t) (= \bm{u}_1-\bm{u}_2)$,
% so $D\cap D'\neq \{0\}$ and $D\neq D'$.
%
\end{lem}

\medskip

\pf
Suppose that $D=D'$.
Then $\bm{u}_j+\alpha\bm{u}_t\in D'$ for all $j=1,2,\ldots,t-1$,
so if
\begin{eqnarray*}
      \bm{u}_j+\alpha\bm{u}_t
  &=& a_1(\bm{u}_1+\beta\bm{u}_t)+\cdots+a_{t-1}(\bm{u}_{t-1}+\beta\bm{u}_t)\\
  &=& a_1 \bm{u}_1           +\cdots+a_j\bm{u}_j +\cdots+a_{t-1}\bm{u}_{t-1}+\biggl(\beta\sum_{\ell=1}^{t-1}a_\ell\biggr)\bm{u}_{t}\,,
\end{eqnarray*}
then
$a_1=\cdots=a_{j-1}=a_{j+1}=\cdots=a_{t-1}=0$, $a_j=1$, and $\alpha=\beta$.
The converse is trivial, so $D=D'$ if and only if $\alpha=\beta$.
Now suppose that $\alpha\neq\beta$.
Let $\bm{v} \in D \cap D'$ and write
\begin{eqnarray*}
  \bm{v} &=& a_1(\bm{u}_1+\alpha\bm{u}_t)+\cdots+a_{t-1}(\bm{u}_{t-1}+\alpha\bm{u}_t)\\
     &=& b_1(\bm{u}_1+\beta \bm{u}_t)+\cdots+b_{t-1}(\bm{u}_{t-1}+\beta \bm{u}_t)
\end{eqnarray*}
for some $a_1,\ldots,a_{t-1},b_1,\ldots,b_{t-1} \in \mathbb{F}_q$.
Then $a_i=b_i$ for all $i=1,\ldots,t-1$, and so
\[
    (a_1+\cdots+a_{t-1})(\alpha-\beta) = 0\,.
\]
We see that $a_1+\cdots+a_{t-1} = 0$, so
\begin{eqnarray*}
  \bm{v} &=& a_1(\bm{u}_1+\alpha\bm{u}_t)+\cdots+a_{t-1}(\bm{u}_{t-1}+\alpha\bm{u}_t)\\
     &=& a_1\bm{u}_1+\cdots+a_{t-1}\bm{u}_{t-1}\,,
\end{eqnarray*}
and we conclude that $\bm{v}\in D_0$.
\qed

%\newpage
\medskip

\begin{thm}
\label{thm:binary}
Let $C$ be a binary $[n,k]$ code with $3 < d^{\perp} < k + 1$.
Then
\[
  \gamma(C) \leq k-d^{\perp}+2\,.
\]
\end{thm}

\pf
Assume that the theorem is false for $C$.
Set $t=d^{\perp}-1 \,(<k)$.
Let $\bm{g}_1,\ldots,\bm{g}_t,\bm{g}_{t+1}$ be $t+1$ linearly independent column vectors of
a generator matrix~$G$ of~$C$.
By Proposition~\ref{prop:summary}, Lemma~\ref{lem:independent},
and the initial assumption,
the $(t-1)$-dimensional subspace $D$ generated by
$\bm{g}_1+\bm{g}_t,\ldots,\bm{g}_{t-1}+\bm{g}_t$ contains a column vector of $G$, say
\begin{eqnarray*}
  \bm{u} &=&a_1(\bm{g}_1+\bm{g}_t)+\cdots+a_{t-1}(\bm{g}_{t-1}+\bm{g}_t)\\
       &=&a_1\bm{g}_1+\cdots+a_{t-1}\bm{g}_{t-1}+\biggl(\sum_{\ell=1}^{t-1} a_\ell\biggr)\bm{g}_t\,.
\end{eqnarray*}
If $a_i=0$ for some $i$,
then $\{\bm{u},\bm{g}_1,\ldots,\bm{g}_{i-1},\bm{g}_{i+1},\ldots,\bm{g}_t\}$ is linearly dependent,
a contradiction.
It therefore follows that $a_i\neq 0$ for all $i$, and so
\[
  \bm{u} = \bm{g}_1+\cdots+\bm{g}_{t-1}+(t-1)\bm{g}_t\:(\mbox{\rm mod}\:\: 2)\,.
\]
If $t$ is odd,
then $\bm{u}=\bm{g}_1+\cdots+\bm{g}_{t-1}$
and so $\{\bm{u},\bm{g}_1,\ldots,\bm{g}_{t-1}\}$ is linearly dependent,
a contradiction.
Therefore, $t$ is even.
Similarly, the $(t-1)$-dimensional subspace $D'$ generated by $\bm{g}_2+\bm{g}_{t+1},\ldots,\bm{g}_{t}+\bm{g}_{t+1}$
contains a column vector $\bm{v}=\bm{g}_2+\cdots+\bm{g}_{t+1}$ of $G$.
Since $\{\bm{g}_1,\ldots,\bm{g}_{t+1}\}$ is linearly independent,
we have that $\bm{u} \neq \bm{g}_{t+1}$ and $\bm{v} \neq \bm{g}_{1}$.
It follows that $\bm{u}+\bm{v}=\bm{g}_1+\bm{g}_{t+1}$ and so $\{\bm{u},\bm{v},\bm{g}_1,\bm{g}_{t+1}\}$ is linearly dependent.
Hence, $d^\perp\leq 4$ and since $d^\perp\geq 4$, we see that $d^\perp = 4$.
However, $t$ is even, so $d^{\perp} = t+1$ is odd, a contradiction.
We conclude that the theorem must be true for $C$.
\qed

\medskip

Note that
if $C$ is a binary $[n,k]$ code with $d^{\perp} = k+1$,
then $C$ is a binary MDS code and so $k = n-1$.
Therefore, by Propositions~\ref{prop:d=3} and \ref{prop:binaryMDS} and Theorem~\ref{thm:binary},
we obtain the following corollary.

\medskip

\begin{cor}
\label{cor:q=2}
Conjecture~{\rm \ref{conj:conjecture}} is true for all binary linear codes.
\end{cor}

\medskip

Let use now consider linear codes over odd fields.

\medskip

\begin{lem}
\label{lem:2 MDS}
{\rm (Theorem 11 on page 326 in \cite{ms77})}\\
If $C$ is a nontrivial $[n, k \geq 3, n-k+1]$ MDS code over $\mathbb{F}_q$ with $q$ odd,
then $n \leq q+k-2$.
\end{lem}

\medskip

\begin{thm}
\label{thm:q=odd}
Let $C$ be an $[n,k]$ code over $\mathbb{F}_q$ with $d^{\perp}> 3$ and $q$ odd.
Then
\[
  \gamma(C) \leq k-d^{\perp}+2\,.
\]
\end{thm}

\medskip

\pf
Assume that the theorem is false for $C$; that is, that $\gamma(C) > k-d^{\perp}+2 = k - (t - 1)$
where $t = d^{\perp} - 1 \; (\geq 3)$.
Assume further, without loss of generality,
that $G=[I_k\: A]$ is a generator matrix for~$C$.
%From the Singleton bound (cf. \cite{ms77}), it follows that $\ell \leq k+1$.
By Proposition~\ref{prop:summary},
we see that our assumptions imply that
each $(t-1)$-dimensional subspace of $\mathbb{F}_q^k$ contains
at least one of the $n$~column vectors of~$G$.
Let $\bm{g}_1,\ldots,\bm{g}_t$ be $t$ column vectors of~$G$
and note that they are linearly independent since $d^{\perp} = t + 1$.
For convenience, write $\mathbb{F}_q = \{\alpha_0 = 0, \alpha_1,\ldots,\alpha_{q-1}\}$.
For each $i = 0,1,\ldots,q-1$, let $D_i$ denote the subspace generated by
$\bm{g}_1+\alpha_i\bm{g}_t,\ldots,\bm{g}_{t-1}+\alpha_i\bm{g}_t$.
From Lemmas~\ref{lem:independent} and~\ref{lem:subspaces},
$D_0,\ldots,D_{q-1}$ are mutually distinct $(t-1)$-dimensional subspaces of $\mathbb{F}_q^k$.
By assumption, each subspace $D_i$ contains a column vector of $G$
\begin{eqnarray*}
  \bm{u}_i &=&a_1(\bm{g}_1+\alpha_i\bm{g}_t)+\cdots+a_{t-1}(\bm{g}_{t-1}+\alpha_i\bm{g}_t)\\
       &=&a_1\bm{g}_1+\cdots+a_{t-1}\bm{g}_{t-1}+\biggl(\alpha_i\sum_{\ell=1}^{t-1} a_\ell\biggr)\bm{g}_t\,.
\end{eqnarray*}
Now consider the matrix
\[
  G'=[\bm{g}_1,\ldots,\bm{g}_t,\bm{u}_1,\ldots,\bm{u}_{q-1}]\,.
\]
Each of the columns in $G'$ are also columns of $G$
and we claim that these columns are distinct.
Clearly, the $t$ columns $\bm{g}_i$ are distinct since they are linearly independent,
so assume that $\bm{u}_i = \bm{g}_j$ for some $i,j$.
Then
\[
  \bm{g}_j = \bm{u}_i = a_1\bm{g}_1+\cdots+a_{t-1}\bm{g}_{t-1}+\biggl(\alpha_i\sum_{\ell=1}^{t-1} a_\ell\biggr)\bm{g}_t\,.
\]
If $j<t$, then $a_j = 1$ and $a_\ell = 0$ for all $\ell\neq j$ and $\sum_{\ell=1}^{t-1} a_\ell = 0$, a contradiction.
Otherwise $j=t$, and $a_1 = \cdots a_{t-1} = 0$ and $\alpha_i\sum_{\ell=1}^{t-1} a_\ell = 1$, also a contradiction.
Hence, no column $\bm{u}_i$ is equal to any column $\bm{g}_j$.
Finally, assume that $\bm{u}_i = \bm{u}_j$ for some distinct $i,j$;
then by Lemma~\ref{lem:subspaces}, $\bm{u}_i\in D_i\cap D_j\subseteq D_0$.
Therefore, $\bm{u}_i$ is linearly dependent on $\bm{g}_1,\dots,\bm{g}_{t-1}$,
a contradiction since any $t$ columns of $G$ are linearly independent.
We conclude that the columns of $G'$ are distinct.

Since these are also columns of $G$,
any $t$ of these columns are linearly independent and, for instance,
the first $t+1$ column vectors are linearly dependent by construction.
Since all columns of $G'$ are contained in the span of the $t$ vectors $\bm{g}_i$,
the dimension is exactly~$t$.
We see that the code $C'$ is an $[t+q-1, t]$ MDS code.

However, $t + q - 1 > q + t - 2$,
so Lemma~\ref{lem:2 MDS} implies that $C'$ is not an MDS code,
a contradiction.
\qed

\medskip

\begin{exmp}
{\rm
Let $C$ be the $[11,5]$ code over $\mathbb{F}_3$ having generator matrix
\[
G=
\left(\begin{array}{ccccccccccc}
1 & 0 & 0 & 0 & 0 & 1 & 2 & 2 & 2 & 1 & 0\\
0 & 1 & 0 & 0 & 0 & 0 & 1 & 2 & 2 & 2 & 1\\
0 & 0 & 1 & 0 & 0 & 2 & 1 & 2 & 0 & 1 & 2\\
0 & 0 & 0 & 1 & 0 & 1 & 1 & 0 & 1 & 1 & 1\\
0 & 0 & 0 & 0 & 1 & 2 & 2 & 2 & 1 & 0 & 1
\end{array}
\right)\,.
\]
Then the dual code $C^{\perp}$ is an $[11,6,5]$ quadratic residue code.
By Magma calculations (cf.~\cite{magma}), we have that
\[
  A_n^{(1)} =   0\,,\quad
  A_n^{(2)} = 330\,,\quad
  A_n^{(3)} = 825\,,\quad
  A_n^{(4)} = 110\,,\quad
  A_n^{(5)} =   1\,,
\]
where $n = 11$.
%If $M_C$ is the vector matroid obtained from $G$,
%then $c(M_C;3)=2\,(=5-5+2)$
Therefore,
$\gamma(C) = 2 = 5 - 5 + 2 = k-d^{\perp}+2$,
so we see that $C$~attains equality in the bound of Theorem~\ref{thm:q=odd}.}
\end{exmp}

\medskip

Now let $C$ be an $[n,k]$ code over $\mathbb{F}_q$ with $d^\perp := d(C^{\perp})$.
It is possible to determine many of the higher weights of $C$ as follows
% the support weight distributions of size $n$ as follows
(cf.~\cite{HeKlMy77,schaathunthesis}).

\medskip

\begin{lem}
\label{lem:klove}
If $k+1-d(C^{\perp}) < r \leq k$, then
\[
  A_n^{(r)}(C) = \sum_{j=0}^{k-r}(-1)^j\genfrac{[}{]}{0pt}{}{k-j}{k-r-j}_q\binom{n}{j}\,.
\]
\end{lem}

\medskip

Using the above lemma,
we can verify Conjecture~\ref{conj:conjecture} for the cases in which $d^\perp = 4$.

\medskip

\begin{lem}
\label{lem:d=4}
Let $C$ be an $[n,k]$ code over $\mathbb{F}_q$ with $d^{\perp}=4$.
Then
\[
  \gamma(C) \leq k-d^{\perp}+2 \;(=k-4+2=k-2)\,.
\]
\end{lem}

\medskip

\pf
The Singleton Bound
implies that $4=d^{\perp}\leq k+1$ and so $k\geq 3$.
By Lemma~\ref{lem:klove},
\begin{eqnarray*}
      A_n^{(k-2)}(C)\!\!
  &=& \sum_{j=0}^{2}(-1)^j\genfrac{[}{]}{0pt}{}{k-j}{2-j}_q\binom{n}{j}\\
  &=& \genfrac{[}{]}{0pt}{}{k}{2}_q\!\!-\genfrac{[}{]}{0pt}{}{k-1}{1}_qn+\genfrac{[}{]}{0pt}{}{k-2}{0}_q\frac{n(n-1)}{2}\\
  &=& \frac{1}{2}\left(n-\biggl(\genfrac{[}{]}{0pt}{}{k-1}{1}_q + \frac{1}{2}\biggr)\right)^2\!\!
    - \frac{1}{2}\left(\genfrac{[}{]}{0pt}{}{k-1}{1}_q + \frac{1}{2}\right)^2 +q^2\genfrac{[}{]}{0pt}{}{k-1}{2}_q\!\! + \genfrac{[}{]}{0pt}{}{k-1}{1}_q\\
%  &=& \frac{1}{2}\left(n-\biggl(\genfrac{[}{]}{0pt}{}{k-1}{1}_q + \frac{1}{2}\biggr)\right)^2\!\!
%    - \frac{q^{k-1}-1}{2(q-1)}\genfrac{[}{]}{0pt}{}{k-1}{1}_q\!\! - \frac{1}{2}\genfrac{[}{]}{0pt}{}{k-1}{1}_q\!\!
%    - \frac{1}{8} + q^2 \genfrac{[}{]}{0pt}{}{k-1}{2}_q\!\! + \genfrac{[}{]}{0pt}{}{k-1}{1}_q\\
%  &=& \frac{1}{2}\left(n-\biggl(\genfrac{[}{]}{0pt}{}{k-1}{1}_q+\frac{1}{2}\biggr)\right)^2\!\!
%    + \frac{(q^{k-1}-q)(q-1)}{2(q^2-1)}\genfrac{[}{]}{0pt}{}{k-1}{1}_q-\frac{1}{8}\\\!\!
  &=& \frac{1}{2}\left(n-\biggl(\genfrac{[}{]}{0pt}{}{k-1}{1}_q+\frac{1}{2}\biggr)\right)^2\!\!
    + \frac{4(q^{k-1}-q)(q^{k-1}-1)-(q^2-1)}{8(q^2-1)}\,.
\end{eqnarray*}
We see that $A_n^{(k-2)}(C) > 0$;
hence, $\gamma(C) \leq k-2 = k - d^\perp + 2$.
\qed

\medskip

By Proposition~\ref{prop:d=3}, Corollary~\ref{cor:q=2},
Theorem~\ref{thm:q=odd}, and Lemma~\ref{lem:d=4},
we have proven that most cases of Conjecture~\ref{conj:conjecture} are true.
The following corollary states this more explicitly.

\medskip

\begin{cor}
\label{cor:mostcases}
Conjecture~{\rm \ref{conj:conjecture}} is true except, possibly,
for linear codes over $\mathbb{F}_q$ where $q=2^m$ for some $m\geq2$ and $d^{\perp}>4$.
\end{cor}

\medskip

Furthermore, note that Conjecture~\ref{conj:conjecture} is true when
the size $q$ of the field $\mathbb{F}_q$ is sufficiently large, namely when $q\geq n$,
since in that case, there is a codeword whose support is $E$,
and so $\gamma(C) = 1$.

\medskip

To conclude this section,
we will now show that Conjecture~\ref{conj:conjecture} is true under conditional circumstances.
The first of these is the event that the following famous conjecture is true (cf.~\cite[p.~265]{pless}).

\medskip

\indent
{\em The MDS Conjecture}
{\rm Suppose that there is a nontrivial $[n,k]$ MDS code over $\mathbb{F}_q$.\\
Then $n \leq q+1$,
except when $q$ is even and $k=3$ or $k=q-1$ in which case $n \leq q+2$.}

\medskip

\begin{thm}
\label{thm:critical-mds}
If the MDS Conjecture is true and $d^\perp\neq q$,
then Conjecture~\ref{conj:conjecture} is true.
\end{thm}

\medskip

\pf
By Corollary~\ref{cor:mostcases},
it suffices to consider a linear code $C$ over $\mathbb{F}_q$
where $q=2^m$ for some $m\geq2$ and $d^{\perp}>4$.
Note that $t = d^\perp - 1 > 3$ and that $t \neq q - 1$.
Assume contradictorily that $\gamma(C) > k-d^{\perp}+2 = k - (t - 1)$;
then by the proof of Theorem~\ref{thm:q=odd},
we may construct a $[t+q-1, t]$ MDS code $C'$.
However, the MDS Conjecture asserts that this is not possible,
a contradiction.
\qed

\medskip

A second conditional circumstance that implies
the validity of Conjecture~\ref{conj:conjecture} is described in the following theorem.

\medskip

\begin{thm}
\label{thm:critical-codes2}
Suppose that there is an $[n,k_0]$ code $C_0$ over $\mathbb{F}_q$ with $d(C_0^{\perp})=\delta\geq 3$.\\
If $C$ is an $[n,k]$ code over $\mathbb{F}_q$ with $d^{\perp} := d(C^\perp) = \delta$ and $k>k_0$,
then
\[
  \gamma(C) \leq k-d^{\perp}+2\,.
\]
\end{thm}

\medskip

\pf
Let $\bm{g}$ be a codeword of $C_0^{\perp}$ with $\textrm{wt}(\bm{g})=\delta$
and extend this codeword to form
a basis $\{\bm{g}=\bm{g}_1,\ldots,\bm{g}_{n-k}\}$ of $C_0^{\perp}$.
For each $i = 1,\ldots,n-k-1$,
let  $C_i$ be the code generated by $\bm{g}_1,\ldots,\bm{g}_{n-k-i}$\,.
Then $C_i$ is an $[n,n-k-i,\delta]$ code
and so $C_i^{\perp}$ is an $[n,k+i]$ linear code with $d((C_i^{\perp})^\perp)=\delta$.
Hence, there is an $[n,k]$ code $C\,(=C_i^\perp)$ over $\mathbb{F}_q$
with $d(C^{\perp})=\delta$ for each $k = k_0,\ldots,n$.

Then let $C$ and $C'$ be an $[n,k]$ code and an $[n,k-1]$ code, respectively,
over $\mathbb{F}_q$ with $k_0 < k \leq n$ and $d(C^\perp)=d((C')^\perp)=\delta$.
By Lemma~\ref{lem:klove},
\begin{eqnarray*}
A_n^{(k-\delta+2)}(C)
&=&\sum_{j=0}^{\delta-2}(-1)^j\genfrac{[}{]}{0pt}{}{k-j}{\delta-2-j}_q\binom{n}{j}\\
&=&\sum_{j=0}^{\delta-2}(-1)^j\left(\genfrac{[}{]}{0pt}{}{k-1-j}{\delta-2-j}_q+q^{k-\delta+2}\genfrac{[}{]}{0pt}{}{k-1-j}{\delta-3-j}_q\right)\binom{n}{j}\\
&=&\sum_{j=0}^{\delta-2}(-1)^j\genfrac{[}{]}{0pt}{}{k-1-j}{\delta-2-j}_q\binom{n}{j}+q^{k-\delta+2}\sum_{j=0}^{\delta-3}(-1)^j\genfrac{[}{]}{0pt}{}{k-1-j}{\delta-3-j}_q\binom{n}{j}\\
&=& A_n^{(k-\delta+1)}(C') + q^{k-\delta+2}A_n^{(k-1-\delta+3)}(C')\,.
\end{eqnarray*}
By Theorem~\ref{thm:kung96-code},
$A_n^{(k-\delta+2)}(C) \geq q^{k-\delta+2}A_n^{(k-1-\delta+3)}(C') > 0$.
Hence, $\gamma(C) \leq k-d^{\perp}+2$.
\qed

%%%%%%%%%%%%%%%%%%%%%%%%%%%%%
%%%%%%%%%%%%%%%%%%%%%%%%%%%%%

\section{A construction of minimal blocks}
\label{sec:construction}

In this section, we present an infinite class of linear codes
that each attains the bound in Conjecture~\ref{conj:conjecture}.

As defined in \cite{kung06,walton82},
a set $M$ of points of the projective geometry $PG(k- 1,q)$
is an $r$-{\em block} over $\mathbb{F}_q$ for some integer~$r$ with $1\leq r \leq k - 1$
if every $(k-r)$-dimensional subspace in $PG(k-1,q)$ contains at least one point in $M$.
If $X$ is a flat in $M$, a {\em tangent} of $X$ is a $(k-r)$-dimensional subspace $U$ in $PG(k-1,q)$ such that
\[
  M \cap U = X\,.
\]
An $r$-block $M$ is {\em minimal} if every point in $M$ has a tangent, and to be {\em tangential} if every proper nonempty flat in $M$ of rank not exceeding $k-r$ has a tangent.

Alternatively, a matroid $M$ is a {\em tangential $r$-block} over $\mathbb{F}_q$ if the following conditions hold:
\begin{enumerate}
  \item[(i)]   $M$ is simple and representable over $\mathbb{F}_q$.
  \item[(ii)]  $p(M;q^r) = 0$.
  \item[(iii)] $p(M/F;q^r) > 0$ whenever $F$ is a proper nonempty flat of $M$.
\end{enumerate}

A construction of minimal blocks from binary vectors is given in \cite{kung06}.

In the following, let $k$ and $m$ be positive integers with $m\leq k$.
Set $K:=\{1,\ldots,k\}$ and let $T \in \binom{K}{m}$.
Also, suppose that $\mathcal{V}$ is a family of $m-1$ distinct points
$\bm{v}_1,\ldots,\bm{v}_{m-1} \in PG(k-1,q)$ with $\textrm{supp}(\bm{v}_i)\cap T = \emptyset$ for each $i=1,\ldots,m-1$.
%\end{eqnarray*}
Define
\begin{align*}
  X^T          & := \{\bm{x} \in PG(k-1,q)\::\: \;\textrm{supp}(\bm{x})\cap T =\emptyset\}\,,\\
%&&Y_{A_1,\ldots,A_{m-1}}^T:=\{\bm{x} \in PG(k-1,2)\::\: |\textrm{supp}(\bm{x})\cap T| =1,\: \textrm{supp}(\bm{x})\setminus T \neq A_i,\: i=1,\ldots,m-1\}\,,\\
  Y_\mathcal{V}^T & := \{\bm{x} \in PG(k-1,q)\::\: |\textrm{supp}(\bm{x})\cap T| = 1\}
    \setminus \bigcup_{j\in T}\bigcup_{\bm{v}_i\in \mathcal{V}}\{\bm{v}_i+\lambda\bm{e}_j \::\: \lambda \in \mathbb{F}_q-\{0\}\}\,,\\[-5mm]
  Z^T          & := \{\bm{x} \in PG(k-1,q)\::\: \;\textrm{supp}(\bm{x}) \in \binom{T}{2}\}\,,\\
  M            & := X^T\cup Y_\mathcal{V}^T \cup Z^T\,,
\end{align*}
where $\bm{e}_i$ denotes the vector in $\mathbb{F}_q^k$ with a $1$ in the $i$th coordinate and $0$'s elsewhere.

\medskip

\begin{thm}
\label{thm:block}
$M$ is a $(k-m)$-block over $\mathbb{F}_q$.
\end{thm}

\medskip

To prove this theorem,
we need the following lemma.

\medskip

\begin{lem}
\label{lem:order}
For any $j \in T$, define
\[
  M_j := \bigl\{\bm{x} \in M \::\:  j \in \textrm{supp}(\bm{x})\bigr\}\,.
\]
Then the following hold:
\begin{enumerate}
\item[{\rm (1)}] $|M_j|=q^{k-m}$.
\item[{\rm (2)}] For any distinct points $\bm{x}, \bm{y} \in M_j$, there exist $\alpha,\beta \in \mathbb{F}_q$ such that $\alpha\bm{x}+\beta\bm{y} \in M$.
\end{enumerate}
\end{lem}

\medskip

\pf
(1) From the definition of $X^T$, $Y_\mathcal{V}^T$, and $Z^T$, we have that
\begin{align*}
  |X^T \cap M_j|          &= 0\,,\\
  |Y_\mathcal{V}^T \cap M_j| &= (q-1)\bigl(|PG(k-m-1,q)|-(m-1)\bigr)+1=q^{k-m}-(q-1)(m-1)\,,\\
  |Z^T \cap M_j|          &= (q-1)(m-1)\,.
\end{align*}
The equation follows.\\
(2) Write ${}^t\bm{x}=(x_1,\ldots,x_k)$, ${}^t\bm{y}=(y_1,\ldots,y_k)$, and $\bm{z} = x_j^{-1}\bm{x}-y_j^{-1}\bm{y}$,
and note that $\bm{x},\bm{y} \notin X^T$ and that $j\notin \textrm{supp}(\bm{z})$.
If $\bm{x},\bm{y} \in Y_\mathcal{A}^T$,
then $\textrm{supp}(\bm{z}) \cap T=\emptyset$
and so $\lambda\bm{z} \in X^T\subseteq M$ for some $\lambda \in \mathbb{F}_q-\{0\}$.
Next, suppose that $\bm{x} \in Y_\mathcal{V}^T$ and $\bm{y} = y_j\bm{e}_j+y_\ell\bm{e}_\ell\in Z^T$
and note that $\bm{x} - \mu'\bm{e}_j\neq \mu\bm{v}_i$ for each $\bm{v}_i\in\mathcal{V}$ and any $\mu,\mu'\in\mathbb{F}_q-\{0\}$.
Then $\textrm{supp}(\bm{z}) \cap T = \{\ell\}$
and  $\bm{z} + y_j^{-1}y_\ell\bm{e}_\ell = x_j^{-1}(\bm{x}-x_j\bm{e}_j) \neq \mu\bm{v}_i$
for any $\bm{v}_i\in\mathcal{V}$ and any $\mu\in\mathbb{F}_q-\{0\}$,
Hence, $\lambda\bm{z} \in Y_\mathcal{V}^T\subseteq M$ for some $\lambda \in \mathbb{F}_q-\{0\}$.
By symmetry,
the same is true if $\bm{x} \in Z^T$ and $\bm{y} \in Y_\mathcal{V}^T$.
Finally, suppose that $\bm{x},\bm{y} \in Z^T$.
Since $\bm{x}$ and $\bm{y}$ are distinct,
the support $\textrm{supp}(\bm{z})$ consists of either one or two elements of~$T$,
and so $\lambda\bm{z}\in Y_\mathcal{V}^T\cup Z^T\subseteq M$ for some $\lambda \in \mathbb{F}_q-\{0\}$.

In each case, $\alpha\bm{x}+\beta\bm{y} = \lambda\bm{z}\in M$
with $\alpha \!=\!  \lambda x_j^{-1}$
and  $\beta  \!=\! -\lambda y_j^{-1}$
for some $\lambda \!\in\! \mathbb{F}_q-\{0\}$.
\qed

\medskip

{\bf Proof of Theorem~\ref{thm:block}. \ }
Suppose that there exists a $(k-m)\times k$ matrix $H = [\bm{a}_1,\ldots,\bm{a}_k]$ over $\mathbb{F}_q$ of rank $k-m$
such that
\[
  M \cap \{\bm{y} \in PG(k-1,q) \::\: H\bm{y}=\bm{0}\} = \emptyset\,.
\]
Choose any $j \in T$.
For any point $\bm{x}={}^{t}(x_1,\ldots,x_k) \in M_j$, $H\bm{x}\neq \bm{0}$ and so it follows that
\begin{eqnarray}
\label{eqn:A}
\bm{a}_j\neq -x_j^{-1}\sum_{\ell\neq j}x_\ell\bm{a}_\ell\,.
\end{eqnarray}
For any distinct points $\bm{x}={}^t(x_1,\ldots,x_k),\bm{y}={}^t(y_1,\ldots,y_k) \in M_j$,
Lemma~\ref{lem:order}~(2) implies that $H(x_j^{-1}\bm{x}-y_j^{-1}\bm{y})\neq \bm{0}$ and so
\[
  -x_j^{-1}\sum_{\ell\neq j}x_\ell\bm{a}_\ell
  \neq
  -y_j^{-1}\sum_{\ell\neq j}y_\ell\bm{a}_\ell\,.
\]
By Lemma~\ref{lem:order}~(1), we have that
\begin{eqnarray}
\label{eqn:B}
\biggl|\biggl\{-x_j^{-1}\sum_{\ell\neq j}x_\ell\bm{a}_\ell \::\: \bm{x}={}^t(x_1,\ldots,x_k) \in M_j\biggr\}\biggr|=|M_j|=q^{k-m}.
\end{eqnarray}
By (\ref{eqn:A}) and (\ref{eqn:B}),
the column vector $\bm{a}_j$ is not in $\mathbb{F}_q^{k-m}$,
a contradiction.
\qed

\medskip

\begin{thm}
\label{thm:minimal}
Let $M$ be the set of points in $PG(k-1,q)$ defined in Theorem {\rm \ref{thm:block}}.\\
If $m \leq q^{k-m-1}$, then $M$ is a minimal $(k-m)$-block over $GF(q)$.
\end{thm}

\medskip

\pf
Without loss of generality, set $T:=\{k-m+1,\ldots,k\}$.
Choose any point $\bm{x} = {}^t(x_1,\ldots,x_k)\in M$.
We first consider the case $\bm{x} \in X^T$.
Set $\ell:=\max\{j \in K\::\: x_j \neq 0\}$ and note that $\ell\leq m-k$.
Consider a $(k-m) \times k$ matrix
\[
  H = \left[\bm{e}_1,\ldots,\bm{e}_{\ell-1},\bm{b},\bm{e}_{\ell+1},\ldots,\bm{e}_{k-m},\bm{y}_{0},\bm{y}_1,\ldots,\bm{y}_{m-1}\right]\,,
\]
where
\[
  \bm{b} =
  \begin{cases}
    \bm{0}                                     & ,\;\textrm{if} \; \textrm{wt}(\bm{x})=1\\\displaystyle
    -x_\ell^{-1}\sum_{j=1}^{\ell-1}x_j\bm{e}_j & ,\;\textrm{otherwise}
  \end{cases}
\]
and $\bm{y}_0=\bm{e}_\ell,\bm{y}_1,\ldots,\bm{y}_{m-1}$ are mutually distinct points in
\[
  \{\bm{z}+\bm{e}_\ell\::\: \bm{z} \in \mathbb{F}_q^k, \: \textrm{supp}(\bm{z}) \cap (T\cup\{\ell\})=\emptyset\}\,.
\]
We note that there always exist these $m$ points whenever $m\leq q^{k-m-1}$.
Let $U$ be the null space in $PG(k-1,q)$ of the matrix~$H$.
Then $\bm{x} \in U$ but $\bm{y} \notin U$ for any $\bm{y} \in M-\{\bm{x}\}$.

Next we consider the case $\bm{x} \in Y_\mathcal{V}^T$.
We may assume without loss of generality that $\textrm{supp}(\bm{x}) \cap T=\{k-m+1\}$
and write ${}^t\bm{v}_i=(v_1^{(i)},\ldots, v_k^{(i)})$ for any point $\bm{v}_i$ in the family $\mathcal{V}$.
Consider a $(k-m) \times k$ matrix
\[
  H = \left[\bm{e}_1,\ldots,\bm{e}_{k-m},\bm{c},\bm{w}_1,\ldots,\bm{w}_{m-1}\right]\,,
\]
where
\begin{align*}
  \bm{c}   &= -x_{k-m+1}^{-1}\sum_{j=1}^{k-m}x_j\bm{e}_j,\\
  \bm{w}_i &= \sum_{j=1}^{k-m}v_j^{(i)}\bm{e}_j\,,\quad\text{for}\:\, i = 1,\ldots,m-1\,.
\end{align*}
Then the null space $U$ of $H$ is a tangent for~$\bm{x}$.

Finally, consider the case in which $\bm{x} \in Z^T$.
Assume without loss of generality that $\textrm{supp}(\bm{x})=\{k-m+1,k-m+2\}$;
then we can construct the null space $U$ by replacing $\bm{c}$ by $-x_{k-m+1}^{-1}x_{k-m+2}\bm{w}_1$
in the above matrix~$H$.
\qed 

\medskip

By definition, $M$ is a minimal $r$-block over $\mathbb{F}_q$
if and only if $\gamma(C) = r+1$ for the linear code $C$ having generator matrix $G$
whose column vectors are all points in $M$ (cf.~\cite[p.~168]{bo92}).

\medskip

\begin{cor}
Let $M$ be the set of points defined in Theorem~{\rm \ref{thm:block}} with $m=2$,
and let $C$ be the linear code over~$\mathbb{F}_q$
whose generator matrix is obtained from~$M$.
Then $C$ attains the bound in Conjecture~{\rm \ref{conj:conjecture}}.
\end{cor}

\medskip

\pf 
From the definition of $M$, we see that $d^{\perp}=3$
since there are three linearly dependent column vectors in~$G$.
Thus,
% we see that
\[\qquad
  k-2+1=k-1=\gamma(C)\leq k-3+2=k-1\,.\qquad\Box
\]

\end{document}